\documentclass[11pt,aps,preprint]{revtex4}
\usepackage[active]{srcltx}
\usepackage[utf8]{inputenc}
\usepackage{latexsym}
\usepackage{amsmath,amssymb,bbm,braket,bm}
\usepackage{graphicx}

\begin{document}

\title{Quantum Spherical Spins with Local Symmetry}

\author{Pedro R. S. Gomes}
\email{prsgomes@bu.edu,pedrorsg@if.usp.br}
\affiliation{Department of Physics, Boston University\\
Boston, MA, 02215, USA}%

\author{P. F. Bienzobaz}
\email{paulafb@bu.edu,paulafb@if.usp.br}
\affiliation{Department of Physics, Boston University\\
Boston, MA, 02215, USA}%

\begin{abstract}

We construct a quantum system of spherical spins with a continuous local symmetry. 
The model is exactly soluble in the thermodynamic limit and exhibits a number of interesting properties. 
We show that the local symmetry is spontaneously broken at finite as well as zero temperatures, 
implying the existence of classical and quantum phase transitions with a nontrivial critical behavior.  
The dynamical generation of gauge fields and the equivalence with  the $CP^{(\mathcal{N}-1)}$ model in 
the limit $\mathcal{N}\rightarrow\infty$  are investigated. The dynamical generation of gauge fields is a consequence of the 
restoration of the local symmetry.

\end{abstract}
\maketitle


\section{Introduction}\label{Int}

The study of lattice models with local  (gauge) symmetry is interesting  as
it brings together several aspects of statistical mechanics and field theories \cite{Kogut,Polyakov,Itzykson}.
Lattice models are a natural regularization for continuum field theory models
in the sense that the lattice spacing acts as a cutoff. On the other hand, the introduction of the lattice spacing changes
the problem to one of statistical mechanics. It becomes then relevant to analyze the phase diagram of the
system and unveil their critical points.

The enhancement of a global symmetry by a local one brings some obstacle concerning the
existence of phase transitions. This is a consequence of the Elitzur's theorem \cite{Elitzur}, which states that
in a theory with local interactions and local gauge symmetry, the
expectation value of non gauge invariant quantities must vanish. In other words, it states that local symmetries cannot be
spontaneously broken.

For concreteness, let us consider the Ising gauge model. There, the Elitzur's theorem implies that the expectation value of the 
order parameter, the magnetization, must be zero since it is not gauge invariant. Thus, the system cannot magnetize and the existence of a phase transition is obscure. Nonetheless, as showed by Wegner \cite{Wegner}, a phase transition occurs
and it is detected not by an order parameter but rather by the behavior of a gauge invariant correlation function.
Moreover, the phase transition in this case can be interpreted as a condensation of topological objects (kinks) \cite{Fradkin}.

Let us analyze the above situation from the energetics perspective.
The crucial point in the Ising gauge model is that, even in the presence of an external field which
breaks the gauge invariance, spin configurations related by a local gauge transformation
differ only by a finite amount of energy. Essentially,
the gauge symmetry makes the expectation value of non invariant quantities, like the order parameter,
insensitive to all degrees of freedom except small neighbors sites (this is in contrast with the usual Ising model
with global symmetry, where all spins cooperate to break the symmetry). This finite number of degrees of freedom
is not strong enough to break the local symmetry and consequently there is no magnetic phase transitions.

That analysis signs that the situation can be drastically different when we have an infinity number of degrees of freedom at
each site or link of the lattice. This occurs for example in systems with internal symmetries, generally
characterized by a number $\mathcal{N}$ of copies of certain variable at each site or link (the same is true
for continuum theories).
If $\mathcal{N}$ tends to infinity, the above analysis is no longer applicable.
For example, consider the Potts model of $q$ states with a local discrete symmetry.
As discussed in \cite{Sondhi}, there is a spontaneous breaking of the
local symmetry for $q\rightarrow\infty$. In addition, a number studies involving the large $\mathcal{N}$ limit in
different systems have been reported showing the spontaneous breaking of
discrete as well as continuous local symmetries \cite{Sondhi,Kugo,Samuel,Samuel1,Green,Vecchia2,Sokal}.

Our intention here is to pursue these ideas through the study of
a quantum system of spherical spins with a continuous local symmetry. Specifically,
we construct a gauge invariant version of the ferromagnetic quantum spherical model with
short-range interactions between first neighbors and explore some consequences. 
We follow an approach in which only the spatial part is discretized, i.e.,
defined on a hypercubic lattice, while the time is maintained as a continuum variable parameterizing the
quantum dynamics of the system. The spin variables, treated as \textquotedblleft matter\textquotedblright  fields (not as gauge fields), are attached to the
sites of the lattice. The auxiliary gauge fields are defined as usual on the links between the sites.
We show that the model exhibits nontrivial classical and quantum phase transitions, corresponding to the
spontaneous breaking of the local symmetry, and determine the critical dimensions.
We also discuss the mechanism of dynamical generation of the gauge fields,
where auxiliary gauge degrees of freedom become dynamical (physical) ones
when the system is quantized. This turns out to be related to the restoration of the local symmetry. As by-product, this analysis unveils some similarities with the $CP^{(\mathcal{N}-1)}$
model which is shown to be concretized  in the large $\mathcal{N}$ limit.

This work is organized as follows. In the Sec. II, we discuss some basics properties of the
quantum spherical model. Sec. III is dedicated to the construction of the model with the local symmetry.
The evaluation of the partition function and the analysis of the critical behavior are subject
of the section IV. In Sec. V, by computing the two point correlation
function of the gauge field, we verify that a Maxwell term for the gauge field
is generated in the continuum limit. The equivalence
with the $CP^{(\mathcal{N}-1)}$ model is shown in Sec. VI. A summary and additional comments
are presented in the Sec. VII. There is also an appendix with some auxiliary calculations.

\section{Quantum spherical model}\label{SecQSM}

Before considering the gauge generalization, let us discuss briefly some basic aspects of the spherical model, starting
with the classical version and following the steps toward its quantization. Since its conception, the classical spherical model \citep{Berlin} has been extensively used in statistical
mechanics as a prototype model to investigate several properties of the critical behavior \citep{Itzykson,Joyce}.
One of its appealing features is that it is a non mean-field model exactly soluble.
Its quantum counterparts \citep{Obermair,Henkel,Nieuwenhuizen,Vojta} share these features,
providing a framework to investigate the presence of quantum fluctuations in a
variety of situations \cite{Tu,Zaleski,Zaleski1}.

The classical model is defined by attaching classical continuous \textquotedblleft spin\textquotedblright variables 
$S_{\bf r}\in\mathbb{R}$ to each site ${\bf r}$ of
a $d$-dimensional hypercubic lattice, interacting according to the Hamiltonian
\begin{equation}
\mathcal{H}_c=\frac12\sum_{{\bf r},{\bf r}'}J_{{\bf r},{\bf r}'}S_{\bf r}S_{{\bf r}'}-h\sum_{\bf r}S_{\bf r},
\label{1.1}
\end{equation}
which is a kind of \textquotedblleft continuous approximation\textquotedblright of the Ising model. The interaction energy
$J_{{\bf r},{\bf r}^{\prime}}$ depends only on the distance between the sites
${\bf r}$ and ${\bf r}'$, $J_{{\bf r},{\bf r}'}\equiv J(|{\bf r}-{\bf r}'|)$, and $h$ is an external field.
The spin variables are subject to the spherical constraint
\begin{equation}
\sum_{\bf r}S_{\bf r}^2=N,
\label{1.2}
\end{equation}
being $N$ the total number of lattice sites. When there is homogeneity in the spins, this condition 
is equivalent to $\langle S_{\bf r}^2\rangle \sim 1$, as in the Ising model. 

In order to quantize the model we have to introduce a dynamics to the system.
This can be done by adding a kinetic term to the Hamiltonian (\ref{1.1}).
If we choose the kinetic term to be  quadratic  in the conjugated momentum to $S_{\bf r}$,
we obtain the following Lagrangian in the absence of the external field
\begin{equation}
L=\frac{1}{2g}\sum_{\bf r}\left(\frac{\partial S_{\bf r}}{\partial t}\right)^2-\frac12\sum_{{\bf r},{\bf r}'}J_{{\bf r},{\bf r}'}S_{\bf r}S_{{\bf r}'},
\label{1.3}
\end{equation}
where $g$ is the quantum coupling.  Note that a term with only one derivative like
$\sum_{\bf r}S_{\bf r}\frac{\partial }{\partial t}S_{\bf r} $ has
no effect since it is just the derivative of the constraint $\sum_{\bf r}S_{\bf r}^2=N$
(this is not the case when the spins are complex, as we shall see).
Now we can proceed with quantization by means of the path integral.
In the imaginary time formalism, we need to pass to the Euclidean imaginary time $\tau=it$, with $\tau\in[0,\beta]$ and
$\beta$ being the inverse of the temperature. Furthermore, the bosonic variables are
required to satisfy the periodic boundary condition $S_{\bf r}(0)=S_{\bf r}(\beta)$.
The partition function is
\begin{equation}
\mathcal{Z}=\int \mathcal{D}S\,\delta\Big{(}\sum_{\bf r}S_{\bf r}^2-N\Big{)}e^{-\int_{0}^{\beta}d\tau{L}_E},
\label{1.4}
\end{equation}
where $L_E$ is the Euclidean form of the Lagrangian (\ref{1.3}).
The integration measure $\mathcal{D}S$ symbolically stands for functional integration over the spins variable of all sites of the lattice,
i.e., $\mathcal{D}S\equiv \prod_{\bf r} \mathcal{D}S_{\bf r}$.
This partition function can be evaluated via the saddle point method,
which becomes exact in the thermodynamic limit.
From the saddle point condition we can determine the critical
behavior of the model and, in particular, check that the model has a nontrivial critical behavior, exhibiting classical and quantum phase transitions \cite{Bienzobaz}.

An interesting aspect of the quantum spherical model is its connection to the $O(\mathcal{N})$ nonlinear sigma model,
involving $\mathcal{N}$ scalar fields $\varphi_a$, $a=1,...,\mathcal{N}$, with the constraint $\sum_a\varphi_a^2=\text{cte}$ and the Lagrangian
\begin{equation}
\mathcal{L}=\frac12\partial_{\mu}\varphi\partial^{\mu}\varphi,
\end{equation}
where we are omitting the sum over $a$.
In fact, the continuum limit  of the quantum spherical model with short-range interactions, where
the Fourier transform of the interaction behaves as $J(q)\sim q^2$, is equivalent to the large $\mathcal{N}$ limit,
$\mathcal{N}\rightarrow\infty$, of the nonlinear sigma model \cite{Vojta,Gomes}.

Now let us consider the case where the spins are complex variables, denoted by $Z_{\bf r}\in\mathbb{C}$  ($\bar{Z}_{\bf r}$
is the complex conjugated).
The spherical constraint becomes $\sum_{\bf r}|Z_{\bf r}|^2=N$ and
the Lagrangian (\ref{1.3}) is generalized to
\begin{equation}
L=\frac{1}{2g}\sum_{\bf r}\frac{\partial \bar{Z}_{\bf r}}{\partial t}\frac{\partial {Z}_{\bf r}}{\partial t}-\frac12\sum_{{\bf r},{\bf r}'}J_{{\bf r},{\bf r}'}(Z_{\bf r}\bar{Z}_{{\bf r}'}+\bar{Z}_{\bf r}{Z}_{{\bf r}'}).
\label{c1.1}
\end{equation}
We see that the Lagrangian as well as the constraint are invariant under global phase
transformations, $Z_{\bf r}\rightarrow e^{i\Lambda}Z_{\bf r}$ and $\bar{Z}_{\bf r}\rightarrow e^{-i\Lambda}\bar{Z}_{\bf r}$,
with $\Lambda$ constant. In addition to its intrinsic interest,
this model is of relevance for the study of superconductivity,
since it can be viewed as a spherical approximation for the Hamiltonian describing a Josephson junction array system
\cite{Kopec,Kopec1,Cha}.

A natural step to further generalize the Lagrangian (\ref{c1.1}) is to
require the invariance under local phase transformations of the spins.
With this, our purpose here is not only to enlarge the present theoretical framework as making wider the range of applications.
In the remaining of this paper we discuss the construction of a Lagrangian with that gauge symmetry
and explore some consequences.

Before closing this section, note that in the case of complex spins
there is the possibility for introducing the
dynamics in a different way \cite{Nieuwenhuizen}, with a kinetic term involving only one time derivative that
was not possible in the case of real spins. The Lagrangian is
\begin{equation}
L=\frac{i}{2g}\sum_{\bf r} \bar{Z}_{\bf r}\frac{\partial {Z}_{\bf r}}{\partial t}-\frac12\sum_{{\bf r},{\bf r}'}J_{{\bf r},{\bf r}'}(Z_{\bf r}\bar{Z}_{{\bf r}'}+\bar{Z}_{\bf r}{Z}_{{\bf r}'}),
\label{c1.2}
\end{equation}
which is also invariant under global phase transformations of the spins.
Higher order time derivatives (higher than two derivatives) generally yield to non unitary quantum evolution.

\section{Including Local Symmetry}\label{Sec3}

The Lagrangian (\ref{c1.1}) is invariant only under global phase transformations of the spins.
We discuss now its generalization to include invariance under local phase transformations in the
case of ferromagnetic short-range interactions, i.e., when the interaction term reduces to
\begin{equation}
\sum_{{\bf r},{\bf r}'}J_{{\bf r},{\bf r}'}(Z_{\bf r}\bar{Z}_{{\bf r}'}+\bar{Z}_{\bf r}{Z}_{{\bf r}'})\rightarrow -J\sum_{<{\bf r},{\bf r}'>}(Z_{\bf r}\bar{Z}_{{\bf r}'}+\bar{Z}_{\bf r}{Z}_{{\bf r}'})=
-J\sum_{{\bf r},I}(Z_{{\bf r}+a\,{\bf e}_I}\bar{Z}_{\bf r}+\bar{Z}_{{\bf r}+a\,{\bf e}_I}{Z}_{\bf r}),
\label{1.5}
\end{equation}
involving only first-neighbors interactions, with $J>0$.
In the last term, $a$ is the lattice spacing and $\{{\bf e}_I\}$, with $I=1,...,d$, is a set of orthogonal unit vectors along all directions,
\begin{equation}
\{{\bf e}_I\}=\{(1,0,0,...,0);(0,1,0,...,0);...;(0,0,0,...,1)\}.
\label{1.6d}
\end{equation}
The local phase transformations are given by
\begin{equation}
Z_{\bf r}\rightarrow e^{i\Lambda_{\bf r}}Z_{\bf r}~~~\text{and}~~~\bar{Z}_{\bf r}\rightarrow e^{-i\Lambda_{\bf r}}\bar{Z}_{\bf r},
\label{1.7a}
\end{equation}
where $\Lambda_{\bf r}$ is an arbitrary real function of the lattice.
The spherical constraint
\begin{equation}
\sum_{\bf r}|Z_{\bf r}|^2=N
\label{1.6}
\end{equation}
is automatically gauge invariant.

In order to motivate the necessary modification of the term (\ref{1.5}), let us rewrite it in a
convenient way. In the continuum limit, $a\rightarrow 0$, we have
\begin{equation}
(Z_{{\bf r}+a\,{\bf e}_I}\bar{Z}_{\bf r}+\bar{Z}_{{\bf r}+a\,{\bf e}_I}{Z}_{\bf r})\approx 2|Z_{\bf r}|^2-
a^2|\partial_I Z_{\bf r}|^2.
\label{1.6b}
\end{equation}
The modulus square term of the spin without derivatives is already invariant. On the
other hand, the term involving derivatives can be turned invariant
by replacing the ordinary derivatives $\partial$ by covariant derivatives $D$.
Thus, for an object $\Phi$ transforming as $\Phi\rightarrow e^{i \Lambda}\Phi$, the covariant derivative of this object
is constructed such that its transformation is $D\Phi\rightarrow e^{i \Lambda}D\Phi$. This guarantees that
any function of the modulus $|D\Phi|$ is gauge invariant. We expect that the right hand side of (\ref{1.6b}) is
generalized as
\begin{equation}
2|Z_{\bf r}|^2-a^2|D_I Z_{\bf r}|^2.
\label{1.6c}
\end{equation}
We need to construct a lattice covariant derivative $D_I$, such that the continuum limit reproduces the usual one.
To this end, let us introduce the complex link variable $U_{{\bf r}I}$, with the transformation law
\begin{equation}
U_{{\bf r}I}\rightarrow e^{i\Lambda_{\bf r}}U_{{\bf r}I}\,e^{-i\Lambda_{{\bf r}+a\,{\bf e}_I}}.
\label{1.7}
\end{equation}
Thus, it is easy to realize that a covariant derivative defined as
\begin{equation}
D_I Z_{\bf r}\equiv U_{{\bf r}I}Z_{{\bf r}+a\,{\bf e}_I}-Z_{\bf r},
\label{1.8}
\end{equation}
has the desired transformation property, i.e., $D_I Z_{\bf r}\rightarrow e^{i\Lambda_{\bf r}}D_I Z_{\bf r}$.
As will be clear in the following, it is convenient to require the links variable to satisfy
\begin{equation}
\bar{U}_{{\bf r}I}U_{{\bf r}I}=1.
\label{1.9}
\end{equation}
Thus we can write $U_{{\bf r}I}$ in terms of a real field $A_{{\bf r}I}$
\begin{equation}
U_{{\bf r}I}=e^{i a A_{{\bf r}I}}.
\label{1.10}
\end{equation}
From the transformation (\ref{1.7}) we see that $A_{{\bf r}I}$
\begin{equation}
A_{{\bf r}I}\rightarrow A_{{\bf r}I} -\frac{1}{a}(\Lambda_{{\bf r}+a\,{\bf e}_I}-\Lambda_{\bf r}),
\label{1.11}
\end{equation}
that, in the continuum limit, is the usual transformation of the gauge field,
$A_{I}\rightarrow A_{I} -\partial_I\Lambda$.
We have to check now that in the continuum limit the lattice covariant derivative (\ref{1.8}) reproduces
the usual one. By taking the limit $a\rightarrow 0$ and expanding $U_{{\bf r}I}=1+i a A_{{\bf r}I}+O(a^2)$, we obtain
the correct result
\begin{equation}
D_I Z_{\bf r}\approx a(\partial_I +i A_{{\bf r}I})Z_{\bf r}+O(a^2).
\label{1.12}
\end{equation}
In this way, the contribution for the Lagrangian of the form
\begin{equation}
(\overline{D_I Z_{\bf r}})( D_I Z_{\bf r})=-\bar{U}_{{\bf r}I}\bar{Z}_{{\bf r}+a\,{\bf e}_I}Z_{\bf r}-
{U}_{{\bf r}I}\bar{Z}_{\bf r}{Z}_{{\bf r}+a\,{\bf e}_I}+|{Z}_{{\bf r}+a\,{\bf e}_I}|^2+
|{Z}_{{\bf r}}|^2,
\label{1.13}
\end{equation}
satisfies the requirements of gauge invariance, Hermiticity, and the correct continuum limit which is consistent with (\ref{1.6c}).
When summed in ${\bf r}$ and $I$, the two last terms become additive constants due the constraint (\ref{1.6}) and can be discarded.

The last step is to analyze the kinetic term. As the time is a continuous variable, the generalization is in terms of usual covariant derivatives
\begin{equation}
\partial_t S_{\bf r}\rightarrow D_t Z_{\bf r}=(\partial_t+iA_{{\bf r}0})Z_{\bf r},
\label{1.14}
\end{equation}
where the $A_{{\bf r}0}$ is the time component of the gauge field, with the transformation $A_{{\bf r}0}\rightarrow A_{{\bf r}0}-\partial_t \Lambda_{\bf r}$.
Of course, the gauge transformation must involve a time-dependent arbitrary function $\Lambda_{\bf r}(t)$.

Collecting all the results, we conclude that the Lagrangian
\begin{eqnarray}
L&=&\frac{1}{2g}\sum_{\bf r} (\overline{D_t Z_{\bf r}})( D_t Z_{\bf r})
-\frac{J}{2}\sum_{{\bf r},I}(\overline{D_I Z_{\bf r}})( D_I Z_{\bf r})\nonumber\\
&=& \frac{1}{2g}\sum_{\bf r} (\overline{D_t Z_{\bf r}})( D_t Z_{\bf r})
+\frac{J}{2}\sum_{{\bf r},I}\bar{U}_{{\bf r}I}\bar{Z}_{{\bf r}+a\,{\bf e}_I}Z_{\bf r}
+\frac{J}{2}\sum_{{\bf r},I}{U}_{{\bf r}I}\bar{Z}_{\bf r}{Z}_{{\bf r}+a\,{\bf e}_I}+\text{cte},
\label{1.15}
\end{eqnarray}
together with the spherical constraint (\ref{1.6}) define a U(1) gauge-invariant generalization of the short-ranged
quantum spherical model.
Note that by taking the limits $A_{{\bf r}0}$ and $A_{{\bf r}I}\rightarrow 0$ or, equivalently, $A_{{\bf r}0}\rightarrow 0$ and
${U}_{{\bf r}I},\bar{U}_{{\bf r}I}\rightarrow 1$,
we recover the model (\ref{c1.1}) for the case of short-range interactions.

As the Lagrangian (\ref{1.15}) does not involve dynamical terms for the gauge fields $A_{{\bf r}0}$ and $A_{{\bf r}I}$,
they can be eliminated by means of their equations of motion. Notice that the fields $A_{{\bf r}0}$ and $A_{{\bf r}I}$
have a very different status in the lattice formulation.
The equation of motion of $A_{{\bf r}0}$ is
\begin{equation}
A_{{\bf r}0}=-\frac{1}{2i |Z_{\bf r}|^2} (\bar{Z}_{\bf r}\overleftrightarrow{\partial_t} Z_{\bf r}),
\label{1.15a}
\end{equation}
where $A\overleftrightarrow{\partial} B\equiv A\partial B-(\partial A) B$, and the equation of motion of $A_{{\bf r}I}$ is
\begin{equation}
A_{{\bf r}I}=\frac{1}{2ia}\text{ln}\left(\frac{\bar{Z}_{{\bf r}+a\,{\bf e}_I}Z_{\bf r}}{\bar{Z_{\bf r}} Z_{{\bf r}+a\,{\bf e}_I}}\right).
\label{1.15b}
\end{equation}
Plugging back these equations into the Lagrangian (\ref{1.15}), we find the clumsy form
\begin{equation}
L=\frac{1}{2g}\sum_{\bf r}\left[ \frac12 |\partial_t Z_{\bf r}|^2+\frac14 \frac{Z_{\bf r}}{\bar{Z}_{\bf r}}(\partial_t \bar{Z}_{\bf r})^2+
\frac14 \frac{\bar{Z}_{\bf r}}{{Z}_{\bf r}}(\partial_t {Z}_{\bf r})^2\right]
+J\sum_{{\bf r},I}|\bar{Z}_{{\bf r}+a\,{\bf e}_I}Z_{\bf r}|,
\label{1.15c}
\end{equation}
in addition to the constraint (\ref{1.6}). It is easy to check that it is invariant under the
local phase transformations (\ref{1.7a}).


\section{Partition Function}

Now let us proceed with the quantization of the model in the form (\ref{1.15}).
The partition function is defined in terms of the path integral
\begin{equation}
\mathcal{Z}=\int \mathcal{D}Z\mathcal{D}\bar{Z}\mathcal{D}A\,\delta\Big{(}\sum_{\bf r}|Z_{\bf r}|^2-N\Big{)}e^{-\int_{0}^{\beta}d\tau{L}_E},
\label{1.16}
\end{equation}
where $L_E$ is the Euclidean form of (\ref{1.15}), i.e., $\tau=it$, with $\tau\in[0,\beta]$.
Here we have included the integration over the gauge fields in the functional integration measure,
$\mathcal{D}A\equiv \prod_{\bf r}\prod_I \mathcal{D}A_{{\bf r}0}\mathcal{D}A_{{\bf r}I}$, besides the complex spin measure
$\mathcal{D}Z\equiv \prod_{\bf r} \mathcal{D}Z_{\bf r}$ and similarly for $\bar{Z}$.
We can take advantage of the thermodynamic limit $N\rightarrow\infty$ to
use the saddle point method to evaluate the partition function.
The strategy is to employ the functional integral representation to the delta functional
\begin{equation}
\delta\big{(}\sum_{\bf r}|Z_{\bf r}|^2-N\big{)}=\int\mathcal{D} \lambda\, e^{{-\int_{0}^{\beta}}d\tau\lambda\big{(}\sum_{\bf r}|Z_{\bf r}|^2-N\big{)}},
\label{1.17}
\end{equation}
which enable us to do the integration over $Z$ and $\bar{Z}$.
We can write the partition function as
\begin{equation}
\mathcal{Z}=\int \mathcal{D}Z\mathcal{D}\bar{Z}\mathcal{D}A\mathcal{D} \lambda\,
\text{exp}\left[+N\int_{0}^{\beta}d\tau \lambda -
\sum_{{\bf r},{\bf r}^{\prime}}\int_{0}^{\beta}d\tau \bar{Z}_{\bf r}M_{{\bf r},{\bf r}^{\prime}}(\partial^2/\partial\tau^2)Z_{{\bf r}^{\prime}}\right],
\label{1.18}
\end{equation}
where the matrix $M$ is defined as
\begin{equation}
M_{{\bf r},{\bf r}^{\prime}}(\partial^2/\partial\tau^2)\equiv \delta_{{\bf r},{\bf r}^{\prime}}\left[-\frac{1}{2g}(\partial_{\tau}+A_{{\bf r} 0})^2+\lambda\right]
-\frac{J}{2}\sum_{I}\bar{U}_{{\bf r}'I}\delta_{{\bf r},{\bf r}'+a{\bf e}_I }-\frac{J}{2}\sum_{I}{U}_{{\bf r}I}\delta_{{\bf r}',{\bf r}+a{\bf e}_I }.
\label{1.19}
\end{equation}
We can integrate out the complex spin fields to obtain
\begin{equation}
\mathcal{Z}=\int \mathcal{D}A\mathcal{D} \lambda\,e^{-NS_{eff}},
\label{1.20}
\end{equation}
with the effective action given by
\begin{equation}
S_{eff}=-\int_{0}^{\beta}d\tau \lambda+\frac{1}{N}\text{Tr}\text{ln} M.
\label{1.21}
\end{equation}
The trace is taken with respect to the matrix elements ${\bf r}$  and ${\bf r}^{\prime}$ as well as with respect to the
derivative operator $\partial_{\tau}$.
In the thermodynamic limit, $N\rightarrow\infty$, the saddle point is determined by the
three conditions
\begin{equation}
\frac{\delta S_{eff}}{\delta\lambda}=\frac{\delta S_{eff}}{\delta A_{{\bf r}0}}=\frac{\delta S_{eff}}{\delta A_{{\bf r}I}}=0.
\label{1.22}
\end{equation}
The saddle point solutions are $\lambda^{\ast}=\text{cte}$ and $A_{{\bf r}0}^{\ast}=A_{{\bf r}I}^{\ast}=0$. The two last conditions
are identically satisfied with these solutions, which is shown in the appendix \ref{appendixA}. The
first condition imposes that
\begin{equation}
1-\frac{1}{N}\sum_{\bf q}\frac{g}{2\omega_{\bf q}}\coth\left(\frac{\beta\omega_{\bf q}}{2}\right)=0,
\label{1.23}
\end{equation}
with $\omega_{\bf q}^2\equiv 2g[\lambda^{\ast}-J\sum_{I}\cos(q_I a)]$, and
we have passed to the Fourier space. In the thermodynamic limit the
sum over ${\bf q}$ must be understood as an integral $\frac{1}{N}\sum_{\bf q}\rightarrow \int d^d{\bf q}$,
with all components of the vector ${\bf q}$ belonging to the first Brillouin zone of the hypercubic lattice, i.e., $-{\pi}/{a}\leq q_I\leq \pi/a$.
Once determined the saddle point we expand the effective action around it up to the quadratic order in the
fields, $S_{eff}[A,\lambda]=S_{eff}^{(0)}+S_{eff}^{(2)}[A,\lambda]$. The zero
order term corresponds to a multiplicative constant in the partition function while the second order term
corresponds to the Gaussian integrations. More specifically,
the quadratic term is decomposed in two parts,
$S_{eff}^{(2)}[A,\lambda]=S_{eff}^{(2)}[\lambda]+S_{eff}^{(2)}[A]$,
such that we evaluate the Gaussian integration over each one of them.
The result is $\mathcal{Z}\propto e^{-NS_{eff}^{(0)}}
e^{-\frac12\text{Tr}\ln \frac{\delta^2S_{eff}}{\delta\lambda\delta\lambda}}
e^{-\frac12\text{Tr}\ln \frac{\delta^2S_{eff}}{\delta A\delta A}}$ evaluated at the saddle point, where
we are omitting the index structure as well as the lattice and time dependence in the functional derivatives.

\subsection{Critical Behavior}

Essentially we have to study  the convergence properties of (\ref{1.23})
near the critical point. With this we will be able to find how the parameters $\beta,g$, and $\lambda^{\ast}$
are related in the proximity of the critical point.
The equation (\ref{1.23}), which determines the critical behavior,  also appears in the case of the usual quantum spherical model. 
The critical point is given by $\lambda_c^{\ast}=J\, \text{max}[\sum_{I}\cos(q_I a)]=Jd$, occurring for the critical value of
the ${\bf q}$ vector ${\bf q}_c=(0,0,...,0)$. We will consider the finite and zero temperature cases separately.

\subsubsection{Finite temperature}

Near the critical point the frequency behaves as
$\omega_{\bf q}^2\sim 2g[\lambda^{\ast}-\lambda_c^{\ast}+\frac{Ja^2}{2}{\bf q}^2+\cdots]$
and the sum  in (\ref{1.23}) is dominated by small values of the momentum.
By using the expansion $\coth x=\frac{1}{x}+\frac{x}{3}+\cdots$, we see that at the critical point the
integral is of the form $\int \frac{d^dq}{q^2}\sim q^{d-2}$, which
converges if $d>2$.  So the lower critical dimension at finite temperatures is given by $d_l=2$.

In the parameter space we will run along a line belonging to a plane with $g=\text{cte}$.
By considering the equation (\ref{1.23}) near the critical point and subtracting this same equation
at the critical point it follows that
\begin{equation}
\frac{1}{N}\sum_q\left(\frac{1}{\beta\omega_{\bf q}^2}-\frac{1}{\beta^c(\omega_{\bf q}^c)^2}\right)+
\frac{1}{N}\sum_{q}\frac{1}{12}\left( \beta\omega_{\bf q}-\beta^c\omega_{\bf q}^c \right)+\cdots=0.
\label{cb1}
\end{equation}
Then, in terms of the temperature $T\equiv 1/\beta$, the
relation between the parameters changes according to the spatial dimensionality of the lattice as
\begin{equation}
(T-T_c)\,\,\,\sim\,\,\,\left\{
\begin{array}
[c]{cc}%
(\lambda^{\ast}-\lambda_c^{\ast})^{\frac{d-2}{2}} & ~~ \left( 2< d<4\right)  \\
(\lambda^{\ast}-\lambda_c^{\ast})\ln(\lambda^{\ast}-\lambda_c^{\ast})  & ~~\left( d=4\right)\\
(\lambda^{\ast}-\lambda_c^{\ast}) &~~\left( d>4\right)
\end{array}
\right..
\end{equation}
Immediately we see that the upper critical dimension is $d_u=4$, that is the critical dimension
separating the nontrivial behavior $(2<d<4 )$  from the mean field behavior $(d>4 )$.

\subsubsection{Zero Temperature}

By taking the zero temperature limit, the equation (\ref{1.23}) becomes
\begin{equation}
1-\frac{1}{N}\sum_{\bf q}\frac{g}{2\omega_{\bf q}}=0.
\label{zt1}
\end{equation}
At the critical point the sum behaves as $\int \frac{d^dq}{q^1}\sim q^{d-1}$
and  converges if $d>1$,  defining the lower critical dimension at zero temperatures as $d_l^0=1$.

In this situation we no longer have the temperature parameter. The distance from the critical
point is characterized now by the difference $g-g_c$ instead of the temperature.  Proceeding as
above, we obtain
\begin{equation}
\frac{1}{N}\sum_{\bf q}\left( \frac{g}{2\omega_{\bf q}}-\frac{g_c}{2\omega_{\bf q}^c}\right)=0.
\label{zt2}
\end{equation}
By studying the behavior of the integrals we obtain
\begin{equation}
(g-g_c)\,\,\,\sim\,\,\,\left\{
\begin{array}
[c]{cc}%
(\lambda^{\ast}-\lambda_c^{\ast})^{\frac{d-1}{2}} & ~~ \left( 1< d<3\right)  \\
(\lambda^{\ast}-\lambda_c^{\ast})\ln(\lambda^{\ast}-\lambda_c^{\ast})  & ~~\left( d=3\right)\\
(\lambda^{\ast}-\lambda_c^{\ast}) &~~\left( d>3\right)
\end{array}
\right.,
\label{zt3}
\end{equation}
giving the upper critical dimension $d_u^0=3$. Thus, the nontrivial critical behavior occurs
for dimension $1<d<3$, while the mean field behavior occurs for $d>3$.

The shift of the critical dimensions in one unit when we compare the cases of zero and finite
temperatures can be understood by considering the way that the dynamics was introduced in the system, which
is characterized by the dynamical critical exponent $z$. We added a kinetic term involving two time derivatives,
corresponding to a factor of $p_0^2$ in the momentum space.
On the other hand, near the critical point the dominating power of momentum
in the frequency $\omega_{\bf q}^2$ is quadratic implying that the scaling
factor between space and time is weighted by $z=1$. According to the general scaling analysis, the
quantum critical point of the model in $d$ dimensions is equivalent to the critical point of the
$d+z$-dimensional model at finite temperature, which means that in our case the critical dimensions
at the finite and zero temperature cases differ by a factor of $z=1$. Moreover,
the equal status of space and time coordinates ($z=1$) is
an evidence of a relativistic structure underlying the model which can manifest in some specific limit.
In fact, this will give rise to a relativistic emergent field theory in the continuum limit, as we will discuss later.

\subsection{Order Parameter}

We show now that it is possible to introduce an order parameter which is not gauge invariant,
characterizing so a spontaneous breaking of the local symmetry.  Indeed, let us define the quantity
\begin{equation}
m\equiv \lim_{h\rightarrow 0}\lim_{N\rightarrow\infty}\left\langle  \frac{1}{N}\sum_{\bf r}(Z_{\bf r}+\bar{Z}_{\bf r})\right\rangle .
\label{op1.1}
\end{equation}
It is important to take the limits in the correct order, since they do not commute. 
For concreteness let us consider the case of finite temperature but the same reasoning goes for the
case of zero temperature (we just need to replace $T$ by $g$ in the arguments). Above the critical temperature, the order parameter vanishes in the
absence of the external field while it acquires a nonzero value below the critical temperature even
in the absence of external field.

We can see this by considering the partition function (\ref{1.16}) in the presence of a constant external field $h$
which breaks the local symmetry,  i.e.,
with the coupling  $h\sum_{\bf r}(Z_{\bf r}+\bar{Z}_{\bf r})$ in the Lagrangian. This will imply an additive term proportional
to $\frac{h^2}{(\lambda^*-\lambda^*_c)}$ in the free energy
$f=\displaystyle\lim_{N\rightarrow\infty }\frac{1}{N}F$ and an additive term proportional to
$\frac{h^2}{(\lambda^*-\lambda^*_c)^2}$ in the saddle point condition (\ref{1.23}).
Thus, the order parameter can be calculated according to
\begin{equation}
m=\frac{\partial f}{\partial h}\sim \frac{h}{(\lambda^*-\lambda^*_c)}.
\label{op1.2}
\end{equation}
Above the critical temperature, as $\lambda^*>\lambda^*_c$, $m\rightarrow 0$
in the absence of the external field, $h\rightarrow 0$.  On the other hand, below the
critical temperature $\lambda^*\rightarrow\lambda^*_c$, yielding to an indetermination in the
order parameter $m$ when $h\rightarrow 0$. In this case, we can use the saddle point equation to settle this indetermination.
Thus, by eliminating the factor $(\lambda^*-\lambda^*_c)$ in the saddle point equation according to (\ref{op1.2}) we get
\begin{equation}
(T-T_c)\,\,\,\sim\,\,\,-m^2+\left\{
\begin{array}
[c]{cc}%
(h/m)^{\frac{d-2}{2}} & ~~ \left(  2<d<4\right)  \\
(h/m)\ln(h/m)  & ~~\left( d=4\right)\\
(h/m) &~~\left( d>4\right)
\end{array}
\right..
\label{op1.3}
\end{equation}
In the case of zero temperature we obtain
\begin{equation}
(g-g_c)\,\,\,\sim\,\,\,-m^2+\left\{
\begin{array}
[c]{cc}%
(h/m)^{\frac{d-1}{2}} & ~~ \left( 1< d<3\right)  \\
(h/m)\ln(h/m)  & ~~\left( d=3\right)\\
(h/m) &~~\left( d>3\right)
\end{array}
\right..
\label{op1.4}
\end{equation}
Finally, by taking the limit $h\rightarrow 0$, the order parameter is non zero below the
critical temperature (or critical coupling), in all dimensions in which the model exhibits a phase transition,
implying the spontaneous breaking of the local symmetry.
We stress again the importance of the specific order of the limits.
First we take the thermodynamic limit, $N\rightarrow\infty$, and then we set the external field to zero, $h\rightarrow 0$.
As discussed in Sec. \ref{Int}, the spontaneous breaking of the local symmetry is the reflex
of an infinite number of degrees of freedom cooperating to form an ordered state.
This infinity number of degrees of freedom is due to
the spherical constraint (\ref{1.6}) which effectively introduces an interaction between all spins
of the lattice. As we will see in Sec. \ref{SecCPN}, this turns out to be equivalent to a system with an infinity number
of degrees of freedom at each point of the spacetime.


\section{Dynamical Generation of the Gauge Fields}

An interesting phenomenon occurring in the quantum spherical model with gauge symmetry is
the dynamical generation of gauge fields. 
This mechanism has been intensively studied in the 
relativistic field theory context, starting with a pioneering work \cite{Bjorken}, which was later 
formulated in terms of path integral in \cite{Eguchi,Bender} (For a more pedagogical treatment, see \cite{Rivers}). 
While classically the gauge fields do not have a true dynamics
and can be eliminated from the Lagrangian, as discussed in Sec. \ref{Sec3}, 
a kinetic term for the gauge fields is generated by quantum corrections implying that the
gauge fields acquire a real dynamics. We will show that this phenomenon is a consequence
of the restoration of the local invariance above the critical point.
To explore this question, we have to consider the quadratic terms in the expansion of the
effective action around the saddle point,
\begin{eqnarray}
S_{eff}^{(2)}[A]&\equiv& \frac{1}{2}\int dt\sum_{{\bf r},{\bf r}'}A_{{\bf r}0}A_{{\bf r}'0}\frac{\delta^{2} S_{eff}}{\delta A_{{\bf r}0}\delta A_{{\bf r}'0}}\Big{|}_{A_{{\bf r}0}=A_{{\bf r}I}=0\atop\lambda=\lambda^*}+\int dt
\sum_I\sum_{{\bf r},{\bf r}'}A_{{\bf r}0}A_{{\bf r}'I}\frac{\delta^{2} S_{eff}}{\delta A_{{\bf r}0}\delta A_{{\bf r}'I}}\Big{|}_{A_{{\bf r}0}=A_{{\bf r}I}=0\atop\lambda=\lambda^*}\nonumber\\&+&\frac12\int dt\sum_{I,J}\sum_{{\bf r},{\bf r}'}A_{{\bf r}I}A_{{\bf r}'J}\frac{\delta^{2} S_{eff}}{\delta A_{{\bf r}I}\delta A_{{\bf r}'J}}\Big{|}_{A_{{\bf r}0}=A_{{\bf r}I}=0\atop\lambda=\lambda^*}.
\label{d1}
\end{eqnarray}
Instead of calculating the derivatives of the effective action in the coordinate space, we
can use the expansion in Feynman diagrams in
the momentum space at zero temperature. Our task is to calculate the two point correlation function of the gauge fields.
Let us first determine the Feynman rules of the model. 
From the free part of the Lagrangian (\ref{1.15}),
\begin{eqnarray}
L_0=\frac{1}{2g}\sum_{\bf r} (\partial_t\overline{ Z}_{\bf r})( \partial_t Z_{\bf r})
+\frac{J}{2}\sum_{{\bf r},I}\bar{Z}_{{\bf r}+a\,{\bf e}_I}Z_{\bf r}
+\frac{J}{2}\sum_{{\bf r},I}\bar{Z}_{\bf r}{Z}_{{\bf r}+a\,{\bf e}_I}-\lambda^{\ast}\sum_{\bf r}\overline{Z_{\bf r}}Z_{\bf r},
\label{d2}
\end{eqnarray}
with the replacement $\lambda\rightarrow \lambda+\lambda^{\ast}$, 
the $Z$-field propagator in the momentum space reads
\begin{equation}
\Delta_{q,q_0}\equiv\frac{i}{\frac{1}{2g}q_0^2-\lambda^{\ast}+J\sum_I\cos(a q_I)}.
\label{d3}
\end{equation}
The finite temperature case is obtained just by replacing $q_0\rightarrow (2\pi n)/\beta$.
The interaction part of the Lagrangian is
\begin{eqnarray}
L_{int}&=&\frac{1}{2g}\sum_{\bf r}\left[ i (\partial_t\overline{Z}_r) Z_{\bf r} A_{{\bf r}0}-
i \overline{Z}_r (\partial_t Z_{\bf r}) A_{{\bf r}0}+A_{{\bf r}0}^2 \overline{Z}_{\bf r}Z_{\bf r}   \right]\nonumber\\&+&
\frac{J}{2}\sum_{{\bf r},I}\left(\overline{U}_{{\bf r}I} -1\right)\bar{Z}_{{\bf r}+a\,{\bf e}_I}Z_{\bf r}
+\frac{J}{2}\sum_{{\bf r},I}\left({U}_{{\bf r}I}-1\right)\bar{Z}_{\bf r}{Z}_{{\bf r}+a\,{\bf e}_I}-
\lambda\left(\sum_{\bf r}\overline{Z}_{\bf r}Z_{\bf r}-N \right).
\label{d4}
\end{eqnarray}
Since some interaction vertices involve the exponential of the $A_{{\bf r}I}$, it is hard to calculate
the correlation functions of the theory. To further proceed, we can consider the model in the continuum limit,
which still captures the essence of the physics we want to discuss, i.e., the arising of a kinetic term for the gauge fields.
Moreover, this limit is interesting since it reveals a relativistic structure underlying the model,
making evident the connection with the field theoretical $CP^{(\mathcal{N}-1)}$ model,
to be discussed in the next section.

\begin{figure}[!h]
\centering
\includegraphics[scale=0.6]{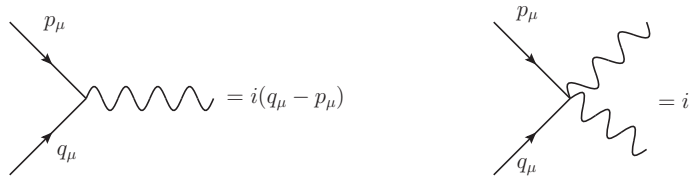}
\caption{Unified vertex factors representing the interaction between the $Z$ and the gauge fields.
The solid and wavy lines represent the $Z$ and gauge fields, respectively.}
\label{VFrelativistic}
\end{figure}

In the continuum limit, the leading contributions come from the following terms in the expansion of
the exponentials in $U_{{\bf r}I}$ and $\bar{U}_{{\bf r}I}$ in powers of $a$:
\begin{eqnarray}
&&\frac{J}{2}\left[\left(\overline{U}_{{\bf r}I} -1\right)\bar{Z}_{{\bf r}+a\,{\bf e}_I}Z_{\bf r}+
\left({U}_{{\bf r}I}-1\right)\bar{Z}_{\bf r}{Z}_{{\bf r}+a\,{\bf e}_I}\right]\nonumber\\&=&
\frac{a^2J}{2} \left[i A_{{\bf r}I}\bar{Z}_{\bf r}(\partial_I Z_{\bf r})-i A_{{\bf r}I}Z_{\bf r}(\partial_I\bar{Z}_{\bf r})
-A_{{\bf r}I}^2\bar{Z}_{\bf r}Z_{\bf r}    \right]+\cdots.
\label{d5}
\end{eqnarray}
From (\ref{d4}) and (\ref{d5})  we see that the vertex factors for the time component
are: $\frac{i}{2g}(q_0-p_0)$ for the trilinear
vertex with two $Z$-lines and one $A_{{\bf r}0}$-line; $\frac{i}{2g}$ for the quartic vertex with
two $Z_{\bf r}$-lines and two $A_{{\bf r}0}$-lines;
and for the spatial lattice components are: $\frac{iJa^2}{2}(q_I-p_I)$ for the trilinear
vertex with two $Z$-lines and one $A_{{\bf r}0}$-line; $\frac{iJ a^2}{2}$ for the quartic vertex with
two $Z_{\bf r}$-lines and two $A_{{\bf r}0}$-lines. The sets $\{q_0,q_I\}$ and $\{p_0,p_I\}$ are the momenta incoming into the vertex along the $Z$-lines.
Observe the emergence of a relativistic structure in the continuum
limit, since the time part becomes of the same type as the spatial part.
At this point, we can take advantage of the relativistic character and use
the "relativistic units" to unify these structures,  where we define $2g\equiv Ja^2/2\equiv1$ such that
the trilinear vertices can be unified into the factor $i(q_{\mu}-p_{\mu})$ and
the quartic vertex into the factor $i$. The Greek index $\mu$ runs over both time and spatial lattice coordinates, $\mu=0,I$.
These two interaction vertices are shown in the Fig. \ref{VFrelativistic}. The propagator in the continuum limit behaves as
\begin{equation}
\Delta_{q,q_0}\sim \frac{i}{q^2-\lambda^{\ast}+Jd}.
\label{d6}
\end{equation}
where we have used the relativistic units and $q^2\equiv q_0^2-\sum_I q_I^2$.
\begin{figure}[!h]
\centering
\includegraphics[scale=0.6]{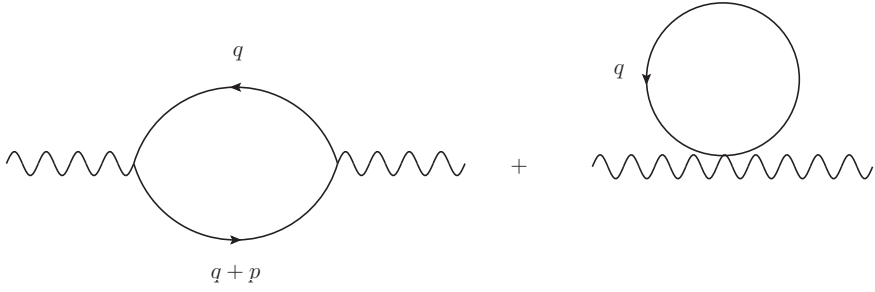}
\caption{Two point correlation function of the gauge field.}
\label{2pf}
\end{figure}

\begin{figure}[!h]
\centering
\includegraphics[scale=0.55]{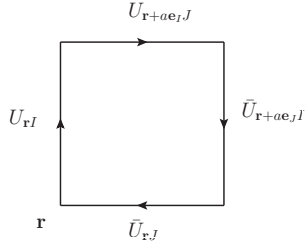}
\caption{A plaquette composed by the product of four link variables. }
\label{plaquette}
\end{figure}

The lowest order terms contributing to the two point correlation function of the gauge fields are the diagrams
of the Fig. \ref{2pf}, with the corresponding expression
\begin{eqnarray}
\Pi_{\mu\nu}(p)&=&\frac{1}{N}\sum_{ q}\frac{(2q+p)_{\mu}(2q+p)_{\nu}}{[(q+p)^2-\lambda^{\ast}+Jd]
[q^2-\lambda^{\ast}+Jd]}-2g_{\mu\nu}\frac{1}{N}\sum_{q}\frac{1}{q^2-\lambda^{\ast}+Jd},
\label{d7}
\end{eqnarray}
where the sum in $q$ involves the integration over $q_0$ and $g_{\mu\nu}=\text{diag}(1,-1,-1,-1,\cdots)$.
The integrations over all components of momentum run over $-\infty$ to $+\infty$ since that,
in the continuum limit, the Brillouin zone extends to the infinity.
Now observe that $p^{\mu}\Pi_{\mu\nu}=p^{\nu}\Pi_{\mu\nu}=0$.
Thus, it follows that, for small momenta (near the critical point), we obtain
\begin{equation}
\Pi_{\mu\nu}(p)\propto \frac{1}{(\lambda^*-\lambda_c^*)^{\frac{4-D}{2}}}(p_{\mu}p_{\nu}-g_{\mu\nu}p^2),
\label{d8}
\end{equation}
remembering that $\lambda^*_c=Jd$.
The result (\ref{d8}) can be straightforwardly checked by using dimensional regularization.
Consider the effective action in the momentum space
\begin{equation}
\frac{1}{N}\sum_{p}A_{\mu} (p)\Pi_{\mu\nu}(p)A_{\nu} (-p).
\label{d9}
\end{equation}
When it is written back in the coordinate space we recognize the Maxwell term
\begin{equation}
S^{(2)}_{eff}[A]\propto \frac{1}{(\lambda^*-\lambda_c^*)^{\frac{4-D}{2}}}\int d^dr dt F_{\mu\nu}F^{\mu\nu},
\label{d10}
\end{equation}
where $D$ is the spacetime dimension, $D=d+1$. This shows that a gauge invariant kinetic term for
the gauge field is induced by quantum corrections, meaning that the auxiliary gauge degrees of
freedom in the classical theory become physical degrees of freedom in the quantized  theory.
We should interpret the Maxwell term above as coming from the lattice gauge invariant action
\begin{equation}
S_{gauge}\propto\int dt\sum_{\bf r}\sum_{I} (F_{0I})^2+
\int dt\sum_{\bf r}\sum_{I,J}(U_{{\bf r}I}U_{{\bf r}+a{\bf e}_I J}\bar{U}_{{\bf r}+a{\bf e}_J I}\bar{U}_{{\bf r}J}-1),
\label{d11}
\end{equation}
in the continuum limit. In this expression,
$F_{0I}\equiv\partial_0 A_{{\bf r}I}-\frac{1}{a}(A_{{\bf r}+a{\bf e}_I 0}-A_{{\bf r}0})$ and the term
involving the product of $U$'s is the usual plaquette contribution, as shown in Fig. \ref{plaquette}.
By taking the continuum limit of (\ref{d11}) we see that the leading contribution is the relativistic Maxwell term
$\int d^Dx F_{\mu\nu}F^{\mu\nu}$.

Let us consider the dimensions in which the model has a nontrivial critical behavior, $1<d<3$. We see that the
coefficient ${1}/{(\lambda^*-\lambda_c^*)^{\frac{3-d}{2}}}$ of (\ref{d10}) is divergent below the
critical coupling $g_c$ since $\lambda^*\rightarrow\lambda^*_c$. This means that the gauge
fields are frozen to their saddle point values $A_{{\bf r}0}=A_{{\bf r}I}=0$ or, in the continuum limit, $A_{\mu}=0$, and 
consequently the local symmetry is broken, $m\neq 0$. Above the critical coupling, on the  other hand,
$\lambda^*>\lambda_c^*$ and the gauge fields propagate according to the Maxwell term
while the symmetry is restored, $m=0$. In conclusion, the dynamical generation of the gauge field is a signal of the
restoration of local symmetry. This in accordance with the results of reference \cite{Kugo}.


\section{Equivalence with $CP^{(\mathcal{N}-1)}$ model}\label{SecCPN}

All the similarities between the continuum limit of the quantum spherical model with gauge symmetry and
the field theoretical $CP^{(\mathcal{N}-1)}$ model become concrete in the large $\mathcal{N}$ limit,
$\mathcal{N}\rightarrow\infty$.
In fact, as we shall discuss in this section, these two models turn out to be equivalent in the specified limits.
To show the equivalence, we compare the partition function of the
$CP^{(\mathcal{N}-1)}$ model with that one of the gauged quantum spherical model.
This is the extension of the equivalence between the quantum spherical model and the nonlinear sigma model
mentioned in the end of the Sec. \ref{SecQSM}.

For completeness, we now review briefly some basic ingredients of the $CP^{(\mathcal{N}-1)}$ model.
The complex projective coset space $CP^{(\mathcal{N}-1)}$ is defined by
\begin{equation}
CP^{(\mathcal{N}-1)}\equiv\frac{SU(\mathcal{N})}{SU(\mathcal{N}-1)\otimes U(1)}
\label{em1}
\end{equation}
and can be parametrized in terms of $\mathcal{N}$ complex fields $\{Z_a\}$ satisfying the local constraint
\begin{equation}
\sum_a\bar{Z}_a Z_a\equiv |Z|^2=\frac{\mathcal{N}}{2f},
\label{em2}
\end{equation}
where $f$ is the coupling constant, such that the points $Z$ and $e^{i\Lambda}Z$ are identified, i.e., $Z\cong e^{i\Lambda} Z$.
This identification is translated into the field theory as a gauge invariance $Z\rightarrow e^{i\Lambda} Z$.
We can immediately write down a gauge invariant Lagrangian for the $Z$-field as
\begin{equation}
\mathcal{L}=(\overline{D_{\mu}Z}) (D^{\mu}Z),
\label{em3}
\end{equation}
with the covariant derivative given by $D_{\mu}=\partial_{\mu}+iA_{\mu}$.
The sum over the number of fields is implicity and the model has both  global $SU(\mathcal{N})$ and local $U(1)$ symmetries.
Classically, $A_{\mu}$ is an auxiliary field and can be completely eliminated by means of its
equation of motion.
The $CP^{(\mathcal{N}-1)}$ model exhibits several interesting properties and
has been intensively studied by a number of authors (see for example \cite{Abdalla} and the references therein.
See also \cite{Polyakov}). In particular, the dynamical generation of gauge fields in this model was initially studied in
\cite{DAdda1,DAdda2,Witten}.

Consider the zero temperature partition function.
We introduce a delta function enforcing the constraint (\ref{em2}) and use
its integral representation to integrate out the complex $Z$-field.
The result is $\mathcal{Z}=\int \mathcal{D}A \mathcal{D}\alpha e^{i {S}_{eff}}$.
The effective action admits an expansion in powers of $1/\mathcal{N}$,
\begin{eqnarray}
S_{eff}=\sum_{n=1}^{\infty}\mathcal{N}^{1-\frac{n}{2}}S^{(n)}.
\label{ex1}
\end{eqnarray}
To make sense of this expansion in the large $\mathcal{N}$ limit we shall impose $S^{(1)}=0$,
which gives us the gap equation
\begin{equation}
\frac{1}{2f}-\int \frac{d^Dk}{(2\pi)^D}\frac{1}{k^2-m^2}=0.
\label{em9}
\end{equation}
By integrating over the $k^0$ component, this equation can be compared with (\ref{zt1})
after an appropriate identification between the parameters. Alternatively, we can write (\ref{em9}) by
taking into account the temperature, which means that we need to replace
\begin{equation}
k^0\rightarrow  \frac{2\pi n}{\beta}~~~\text{and}~~~\int dk^0 \rightarrow \frac{1}{\beta}\sum_n.
\end{equation}
After performing the sum over the integers $n$ we recover (\ref{1.23}).
Finally, when we take the limit $\mathcal{N}\rightarrow\infty$, only the quadratic terms will remain in (\ref{ex1}),
$\displaystyle\lim_{\mathcal{N}\rightarrow\infty}S_ {eff}=S^{(2)}[\lambda,A]=S^{(2)}[\lambda]+S^{(2)}[A]$.
The quadratic part in the gauge fields is
\begin{equation}
S^{(2)}[A]=\int \frac{d^Dp}{(2\pi)^D}
A_{\mu}(p)\Pi^{\mu\nu}(p)A_{\nu}(-p),
\label{em10}
\end{equation}
with $\Pi^{\mu\nu}(p)$ as in (\ref{d7}),  showing that the gauge contribution to the effective action
coincides with that one of the quantum spherical model.
We still have to discuss the equivalence in the $\lambda$-sector.
The fundamental difference in this sector is the dependence of the Lagrange multipliers enforcing
the constraints in these theories. In the case of the quantum spherical model, due
to the nonlocal nature of the constraint, the Lagrange multiplier depends only on the time.
On the other hand, in the $CP^{(\mathcal{N}-1)}$ model, the local character of the
constraint implies that the Lagrange multiplier depends on the time as well as on the spatial
coordinates. So the equivalence between the partition functions is reached
after we integrate out the respective $\lambda$-fields with the results evaluated at the saddle point,
$\lambda=\text{cte}$ and $A_{\mu}=0$. As the functional dependence is the same in both cases
(with appropriate comparison in the Euclidean or Minkowski space) we conclude that
the both partition functions become the same.
To sum up, the continuum limit of the partition function of the quantum spherical model with
the gauge symmetry is equivalent to that one of the $CP^{(\mathcal{N}-1)}$ in the limit
$\mathcal{N}\rightarrow\infty$.


\section{Summary}

In this work we have constructed a version of the quantum spherical model
which is invariant under local phase transformations of the complex spins, aiming to
investigate the existence of phase transitions. As we stressed, this is a delicate question
in view of the Elitzur's theorem.  The model was exactly solved in thermodynamic limit by means of the
saddle point method.
By analyzing the properties of the saddle point equation we showed that the model indeed exhibits
a nontrivial (non mean-field) phase transition in both cases of zero and finite temperatures,
corresponding to a spontaneous breaking of the local symmetry characterized by a non invariant order parameter.
This is understood due to an infinite number of degrees of freedom cooperating to form an
ordered state even in the theory with gauge invariance.

By computing the two point correlation function for the gauge fields in the
continuum limit we verified that a Maxwell-type kinetic term is generated for the
gauge field. Moreover, we concluded that the appearance of a dynamical gauge field
is associated to the restoration of the gauge invariance above the critical coupling.
The model was also shown to be equivalent to the limit $\mathcal{N}\rightarrow\infty$ of the relativistic
$CP^{(\mathcal{N}-1)}$ model by comparing the quantum effective actions in both cases.

As a lateral remark, we point the difficulty in implementing the local symmetry in the model (\ref{c1.2}) following
the lines described in the Sec. \ref{Sec3}. This is so because
the classical equation of motion for the auxiliary field $A_{0{\bf r}}$ implies
$\bar{Z}_{\bf r}{Z}_{\bf r}=0$, which seems to be an inconsistency.

Further investigations of the gauge invariant quantum spherical model naturally asks for applications.
At least as an approximation for certain physical problems, it can be used as a prototype model.
In addition, the inclusion of different types of interactions, for example,
second neighbors in a competing way (anti-ferro) could produce a richer phase structure
including Lifishitz points. It is tempting to seek for a field theory equivalence in this case.
Equally interesting is trying to formulate the problem in lattices with different geometries.

\section{Acknowledgments}

The authors thank Marcelo Gomes and Christopher Mudry for carefully reading the manuscript, very useful discussions
and the criticism. We also thank Claudio Chamon for very interesting discussions.
This work was supported by  Funda\c{c}\~ao de Amparo a Pesquisa do Estado de S\~ao Paulo (FAPESP).

\appendix

\section{Saddle point conditions}\label{appendixA}

In this appendix we want to show that $A_{{\bf r}0}=A_{{\bf r}I}=0$ are the solutions of the saddle point conditions
\begin{equation}
\frac{\delta S_{eff}}{\delta A_{{\bf r}0}}=\frac{\delta S_{eff}}{\delta A_{{\bf r}I}}=0.
\label{a1}
\end{equation}
Let us start with the first condition. We will use the identity $\delta \text{Tr} \text{ln} A=\text{Tr}A^{-1}\delta A $.
Employing a coordinate basis to take the trace, it follows that
\begin{equation}
\frac{\delta M_{{\bf r},{\bf r}'}}{\delta A_{\tilde{{\bf r}}0}(\tilde{\tau})}\Big{|}_{A_{{\bf r}0}=A_{{\bf r}I}=0}=
-2\delta_{{\bf r},{\bf r}'}\delta_{{\bf r},\tilde{\bf r}}\partial_{\tau}\delta(\tau-\tilde{\tau}).
\label{a2}
\end{equation}
Thus we have
\begin{eqnarray}
\frac{\delta S_{eff}}{\delta A_{\tilde{{\bf r}}0}(\tilde{\tau})}\Big{|}_{A_{{\bf r}0}=A_{{\bf r}I}=0}&=&
\int_{0}^{\beta}d\tau\langle \tau |\sum_{{\bf r},{\bf r}'} (M^{-1})_{{\bf r},{\bf r}'} \frac{\delta M_{{\bf r}',{\bf r}}}{\delta A_{\tilde{{\bf r}}0}(\tilde{\tau})}|\tau\rangle \Big{|}_{A_{{\bf r}0}=A_{{\bf r}I}=0}\nonumber\\
&=&-2\int_{0}^{\beta}d\tau\langle \tau |\sum_{{\bf r},{\bf r}'} (M^{-1})_{{\bf r},{\bf r}'} \delta_{{\bf r},{\bf r}'}\delta_{{\bf r},\tilde{\bf r}}\partial_{\tau}\delta(\tau-\tilde{\tau})|\tau\rangle. 
\label{a3}
\end{eqnarray}
To see that this expression vanishes, we introduce a basis $|n\rangle$ in which the operator $\partial_{\tau}$ is diagonal, i.e.,
$\frac{\partial}{\partial\tau}|n\rangle=i\omega_n|n\rangle$, with $\omega_n=2n\pi/\beta$, $n\in \text{Z}$. This is so
due to the finite interval $[0,\beta]$ of the variable $\tau$.
The inverse matrix $M^{-1}$ is a function of $\partial^2$, which will become proportional to $n^2$ in the $|n\rangle$ basis.
On the other hand, we will have a term proportional to $n$ in the numerator due to the linear derivative term $\partial_{\tau}$.
Thus we see that when summed over all integers $n$ this relation will vanish.

Now let us consider the second condition in (\ref{a1}). We need to calculate
\begin{equation}
\frac{\delta M_{{\bf r},{\bf r}'}}{\delta A_{\tilde{{\bf r}}\tilde{I}}(\tilde{\tau})}\Big{|}_{A_{{\bf r}0}=A_{{\bf r}I}=0}=
\sum_I\left[(-ia) \frac{J}{2}\delta_{{\bf r}',\tilde{\bf r}}\delta(\tau-\tilde{\tau})\delta_{I,\tilde{I}}\delta_{{\bf r},{\bf r}'+a{\bf e}_I }+
(ia) \frac{J}{2}\delta_{{\bf r},\tilde{\bf r}}\delta(\tau-\tilde{\tau})\delta_{I,\tilde{I}}\delta_{{\bf r}',{\bf r}+a{\bf e}_I }\right].
\label{a4}
\end{equation}
With this, we obtain
\begin{eqnarray}
\frac{\delta S_{eff}}{\delta A_{\tilde{{\bf r}}\tilde{I}}(\tilde{\tau})}\Big{|}_{A_{{\bf r}0}=A_{{\bf r}I}=0}&=&
\int_{0}^{\beta}d\tau\langle \tau |\sum_{{\bf r},{\bf r}',I} (M^{-1})_{{\bf r},{\bf r}'} \frac{\delta M_{{\bf r}',{\bf r}}}{\delta A_{\tilde{{\bf r}}\tilde{I}}(\tilde{\tau})}|\tau\rangle \Big{|}_{A_{{\bf r}0}=A_{{\bf r}I}=0}\label{a5}\\
&=&(-ia)\frac{J}{2}\int_{0}^{\beta}d\tau\langle \tau |\sum_{{\bf r},{\bf r}',I} (M^{-1})_{{\bf r},{\bf r}'}
\left(\delta_{{\bf r}',\tilde{\bf r}}\delta_{{\bf r},{\bf r}'+a{\bf e}_I }-
\delta_{{\bf r},\tilde{\bf r}}\delta_{{\bf r}',{\bf r}+a{\bf e}_I }\right)
\delta(\tau-\tilde{\tau})\delta_{I,\tilde{I}}|\tau\rangle. \nonumber
\end{eqnarray}
We can see that this expression vanishes by noting that the inverse matrix $(M^{-1})_{{\bf r},{\bf r}'}\Big{|}_{A_{{\bf r}0}=A_{{\bf r}I}=0}$
is a symmetric matrix since $M_{{\bf r},{\bf r}'}\Big{|}_{A_{{\bf r}0}=A_{{\bf r}I}=0}$ is symmetric.
On the other hand, the term $(\delta_{{\bf r}',\tilde{\bf r}}\delta_{{\bf r},{\bf r}'+a{\bf e}_I }-
\delta_{{\bf r},\tilde{\bf r}}\delta_{{\bf r}',{\bf r}+a{\bf e}_I })$ is anti-symmetric in the indices ${\bf r}, {\bf r}'$,
yielding to the desired result.


\end{document}